LA-UR-12-22199

Title: DARHT Axis-I Diode Simulations II: Geometrical Scaling

Author(s): Ekdahl, Carl A. Jr.

Intended for: Report

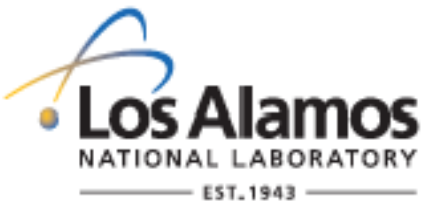

# DARHT Axis-I Diode Simulations II: Geometrical Scaling

*Carl Ekdahl*

## I. INTRODUCTION

Flash radiography of large hydrodynamic experiments driven by high explosives is a venerable diagnostic technique in use at many laboratories. Many of the largest hydrodynamic experiments study mockups of nuclear weapons, and are often called hydrotests for short. The dual-axis radiography for hydrodynamic testing (DARHT) facility uses two electron linear-induction accelerators (LIA) to produce the radiographic source spots for perpendicular views of a hydrotest. The first of these LIAs produces a single pulse, with a fixed ~60-ns pulsewidth. The second axis LIA produces as many as four pulses within 1.6-µs, with variable pulsewidths and separation.

There are a wide variety of hydrotest geometries, each with a unique radiographic requirement, so there is a need to adjust the radiographic dose for the best images. This can be accomplished on the second axis by simply adjusting the pulsewidths, but is more problematic on the first axis. Changing the beam energy or introducing radiation attenuation also changes the spectrum, which is undesirable. Moreover, using radiation attenuation introduces significant blur, increasing the effective spot size. The dose can also be adjusted by changing the beam kinetic energy. This is a very sensitive method, because the dose scales as the ~2.8 power of the energy, but it would require retuning the accelerator. This leaves manipulating the beam current as the best means for adjusting the dose, and one way to do this is to change the size of the cathode. This method has been proposed, and is being tested [1]. This article describes simulations undertaken to develop scaling laws for use as design tools in changing the Axis-1 beam current by changing the cathode size.

## II. PHYSICAL CONSIDERATIONS

### *A. Background*

The standard configuration of the DARHT Axis-I diode features a 5.08-cm diameter velvet emitter mounted in the flat surface of the cathode shroud. The surface of the velvet is slightly recessed. This configuration produces a 1.75 kA beam when a 3.8-MV pulse is applied to the anode-cathode (AK) gap [2]. Velvet cold-cathode emitters produce a plasma surface by the explosive emission process from which a space-charge limited current can be drawn [3,4,5].

Space-charge limited flow of electrons can result from any source of electrons – thermionic, field emission, plasma extraction, photo-emission, or Compton scattering. The maximum current that can be drawn from any of these sources is limited by the space charge of the resulting beam. When the current is space-charge limited, excess electrons from the source are reflected back to the emission surface by the space-charge potential well. Although Child was the first to derive the dependence of space-charge limited ion current on applied voltage [6], Langmuir was the first to derive the space-charge limit for electrons[7], and it was Langmuir who accounted for reflected thermal electrons [8]. Child and Langmuir both derived the familiar $J \propto V_{AK}^{3/2}$ law for infinite planar diodes. Much later Jory and Trivelpiece extended this work into the relativistic range of accelerated electron velocity [9], also finding that the space-charge limited current density in a relativistic diode is directly proportional to a function of AK voltage alone, albeit a more complicated function than the simple $V_{AK}^{3/2}$ of Child and Langmuir.

Finally, Langmuir showed that the dependence of non-relativistic current on AK voltage is independent of the diode geometry, so one only needs the planar diode solution to establish the I-V scaling law [7,10] for any geometry. That is, for all non-relativistic diodes $J \propto V^{3/2}$, with a proportionality constant determined by the specific geometry. This result can be extended into the relativistic diode regime; all relativistic-diode geometries have the same Jory-Trivelpiece scaling of current as a function of AK voltage, regardless of shape.

## *B. DARHT Axis-I Diode*

At first glance, it might seem that the current from the DARHT flat cathode should scale as the area of the cathode. That is, constant current density as in the Child-Langmuir [6,7] or Jory-Trivelpiece [9] laws. But this would be wrong, because the Axis-1 diode is far from the infinite planar diode geometry of these theories. Furthermore, neither of these theories deals with a beam entering a solenoidal magnetic field within the diode gap. Moreover, unless the cathode is slightly recessed, the emission at the cathode edge is higher than at the center because the potential at the center is depressed by the beam space charge. This effect produces a non-uniform, hollowed-out beam, with the higher current density at the edge attributed to "edge-emission." Therefore, one approach to developing accurate scaling laws is through simulations that account for the finite diode geometry, for the external magnetic field, and for the Pierce-like effect [11,12] of recessed cathode production of a beam with relatively uniform current density. The alternative is a large number of experiments to develop an empirical scaling with cathode diameter and depth of recess. On the other hand, only a limited number of careful experiments should be required to validate scaling based on these simulations.

# III. SIMULATIONS

All simulations were performed in 2-D using the TRICOMP suite of codes[13]. The applied electric and magnetic fields were simulated using finite-element methods based on a conformal triangular mesh model of the DARHT Axis-1 diode [14,15]. The TRAK electron-gun simulator in this suite was then used to simulate the space-charge limited beam current.

## A. Applied Electric Field

The electrostatic field solution is based on an accurate model of the DARHT Axis-1 diode and insulator, which was originally developed to inform insulator repairs. I reduced this to more tractable model by eliminating unnecessary details (insulators, grading rings, etc). I also rezoned the mesh to provide more detail in the region of cathode emission, and added the details of the velvet cathode. Figure 1 shows the original model with the region of the refined model shown by heavy dashed lines.

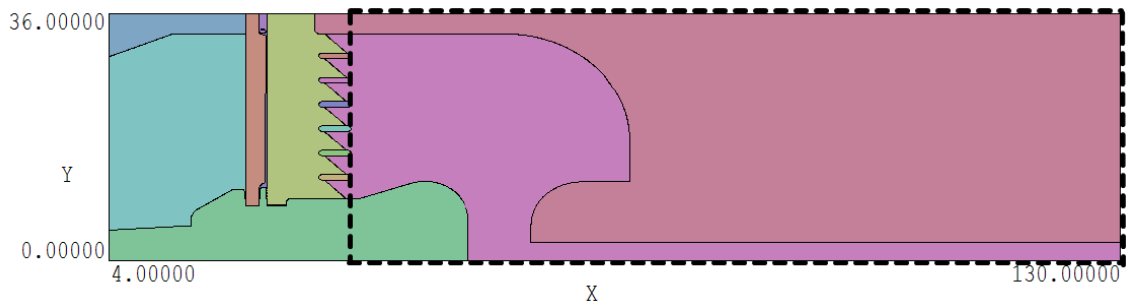

Figure 1: Full model of the DARHT Axis-1 diode and insulator region. The refined model region is outlined. Dimensions are in inches.

To simulate the electric field with no electron beam a negative potential was applied to the center conductor, reported herein as the AK voltage (unsigned). All of the simulations in this article were performed with the nominal Axis-I AK voltage (3.8 MV). The results can be scaled to other AK voltages using the scaling law $I = 0.271 V_{AK}^{1.3898}$ derived from simulations in Ref. [2], because of the independence of geometrical and voltage scaling discussed above and in Ref. [7,8,10]. The resulting electrostatic potentials with no beam are shown in Fig. 2. The potential solutions for the two models (full geometry and refined geometry) differ by less than 0.1% in the AK gap, so the refined model yields acceptable results with a significant savings in computational memory requirements and time.

Figure 3 shows a blowup of the equipotentials in the emitter region. Recessing the surface of the emitter reduces the surface electric field at the center to 185 kV/cm from the 208 kV/cm on the flat face of the shroud when $V_{AK}$=3.8 MV.

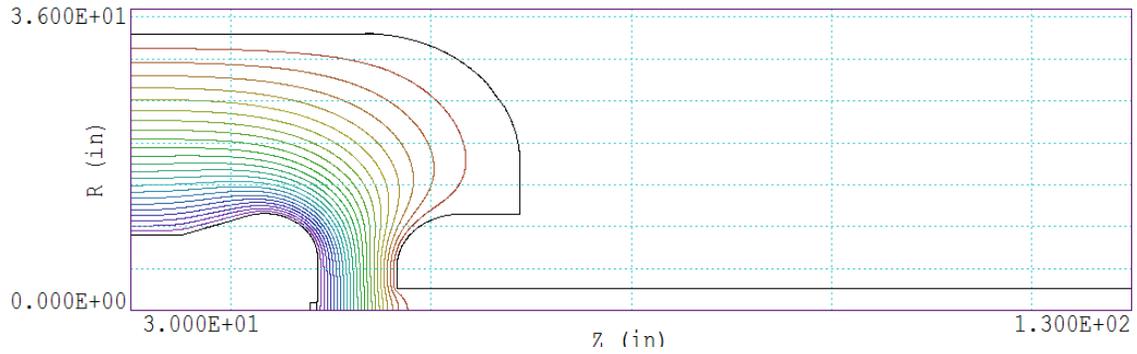

Figure 2: Electrostatic potential in the DARHT Axis-1 diode with no beam present, and $V_{AK}$=3.8 MV.

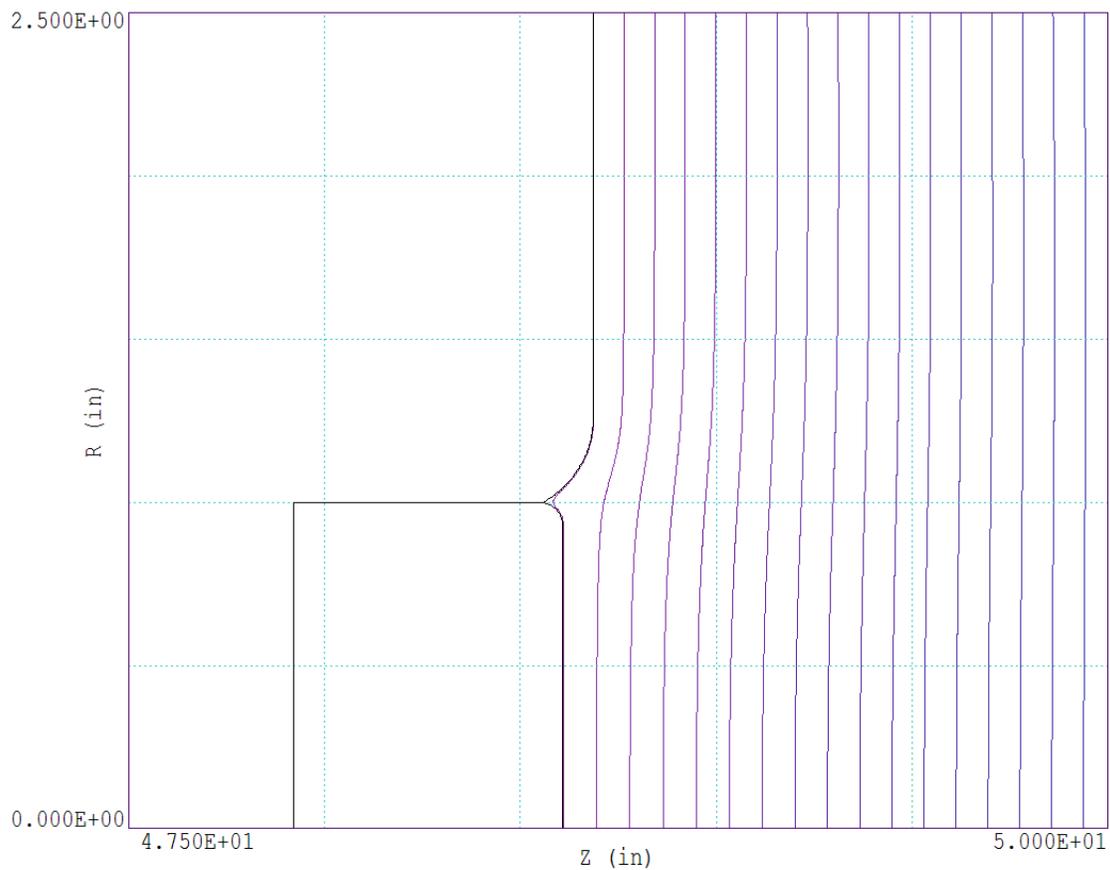

Figure 3: Close-up of the velvet emitter region with no beam. For this particular simulation of a 2-inch diameter cathode the emission surface is recessed 2.0 mm from the flat surface of the shroud. With an diode voltage $V_{AK}$=3.8 MV the field at the surface of the emitter is only 185 kV/cm, compared with 208 kV/cm at the surface of the shroud.

## B. External magnetic field

The magnetic field map was obtained from the PERMAG code [13] by modeling the bucking coil and anode solenoid as ideal sheet solenoids having the dimensions and locations specified by the XTR envelope code used for tuning Axis-1 [2]. The XTR model is a best fit to experimental field measurements. The base-case field was simulated for 100 A energizing the anode solenoid and -0.1451 A energizing the bucking coil. (The ratio of these two currents, $k_{buck}$=0.1451, was determined by inspection of current read-backs from several days of operation.) Figure 4 shows a comparison of the axial magnetic field on axis calculated with the PERMAG simulation and the magnetic field calculated with the XTR model. These simulations give nearly identical results, and since the XTR models are based on field measurements, the magnetic field in these TRAK simulations is presumably in agreement with reality. The field for other anode settings (with fixed $k_{buck}$ =0.1451) is simply obtained by scaling the base-case 100-A solution.

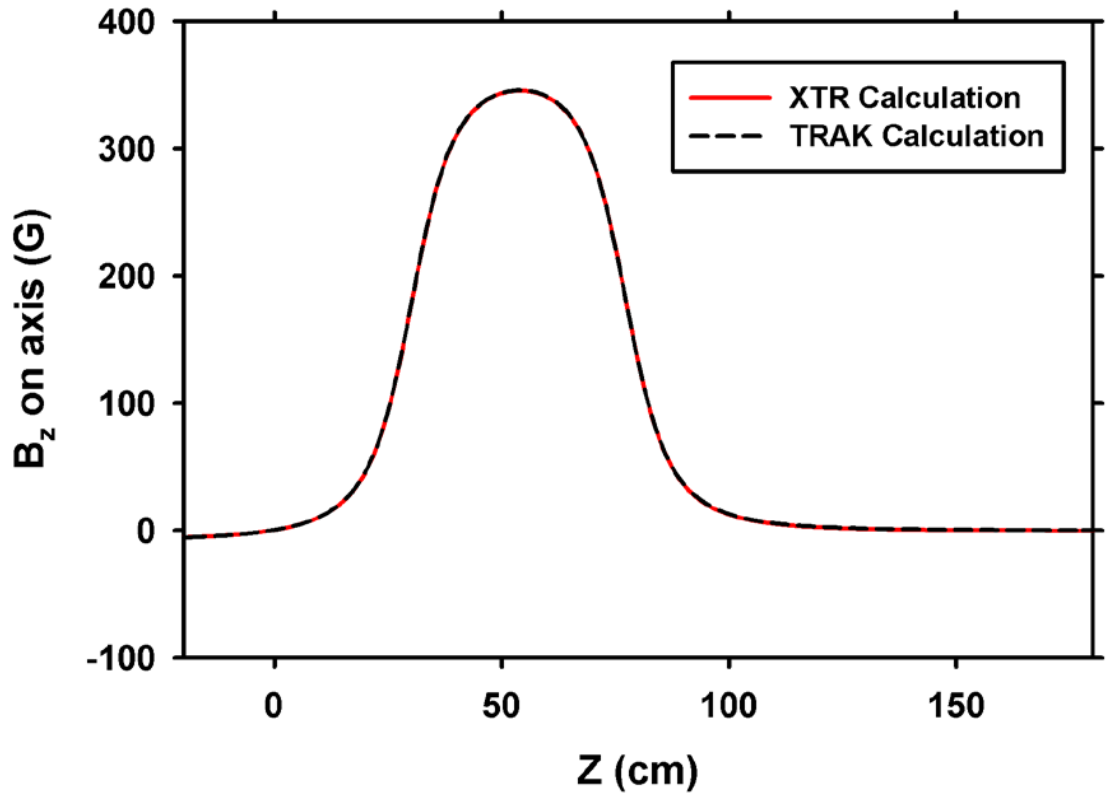

Figure 4: Axial magnetic field calculated on axis by XTR (red, solid curve) and by PERMAG (black, dashed curve) for a 100-Ampere driving current on the anode solenoid, and $k_{buck}$=0.1451.

## C. Electron Beam and Space-Charge Field

Using the applied field solutions as input, the TRAK electron-gun code [13] was then used to simulate the space-charge limited beam produced in the diode. TRAK self-consistently simulates the beam and the beam-generated electric and magnetic fields. The total electric field calculated includes both the applied field and that of the beam space-charge. TRAK simulates space-charge limited emission of electrons, presumably from the plasma formed on the velvet cathode [5]. Emission is calculated self-consistently by iteratively increasing the current from emission elements until the total electric field at the emission surface is zero. This is the Child condition [6], and it actually occurs at the surface where excess emitted electrons are reflected, rather than exactly at the emitter, as explained by Langmuir [8,10]. As also shown by Langmuir, this surface is only slightly displaced from the emitter itself.

The iterative algorithm in TRAK for calculating the space-charge limit rapidly converges in about~7 iterations, and then varies only slightly with further iterations, with the rms variation dependent on the problem setup (mesh size, geometry, etc.). To reduce the uncertainty in the result, I ran all problems for a total of 20 iterations, and averaged the results of the last 14. This gave results for all of these setups with standard deviations in the range of 1-2%, which is adequate for this geometric-scaling analysis.

## III. GEOMETRICAL SCALING

### A. Ideal Flat Cathode

As mentioned above, high-current, flat cathodes produce beams with a lower current density at the center than at the edge. This is because the field is lower at the beam center than at the edge due to the beam space charge. Indeed, it was to overcome this effect that J. R. Pierce developed electron-gun designs with conical electrodes to flatten the equipotentials in the beam, which would otherwise be curved by space charge [11,12]. This effect can be clearly seen in my simulations of an ideal flat emission surface congruent with the flat shroud surface. Figure 5 shows the space-charge limited beam extracted from a flat emission surface that is simply defined as a 5.08-cm (2-inch )diameter emitting patch of with the Axis-I shroud. The equipotential surfaces due to the beam space charge in addition to the applied AK voltage are also shown in this figure.

The equipotentials in Fig. 5 clearly show the space-charge depression on axis near the emitter. The reduced electric field in the center extracts a lower current density, producing a beam with a higher current density at the edge than at the center. This effect, clearly seen in a plot of current density across the emission face (Fig. 6), is often called "edge emission," even though there is no field enhancement by a corner at the edge for this ideal case. It should be noted that in the idealized geometries considered by Langmuir and others (infinite planar, coaxial cylinders, concentric spheres), the field is uniform over the emission surface, so the emission is also uniform, which is significantly different than the case of localized emission surfaces analyzed here.

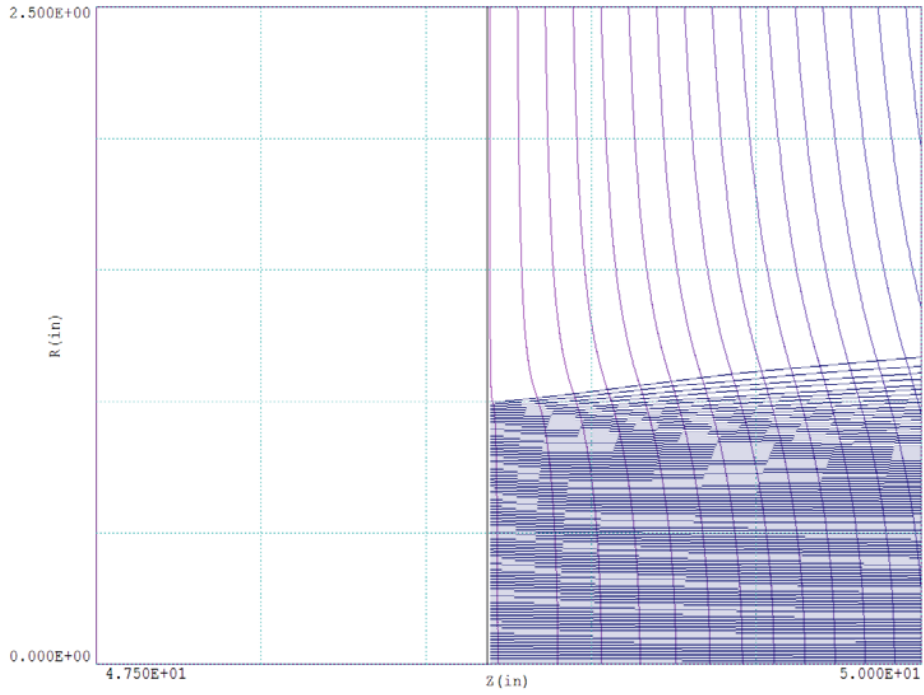

Figure 5: Electron beam extracted from a 2-inch diameter, space-charge limited emitter that is flush with the flat surface of the Axis-I cathode shroud. Simulation of this geometry produced 2.1 kA with $V_{AK}$=3.8 MV. (N. B.,The density of rays is not equivalent to current density because the rays do not
carry

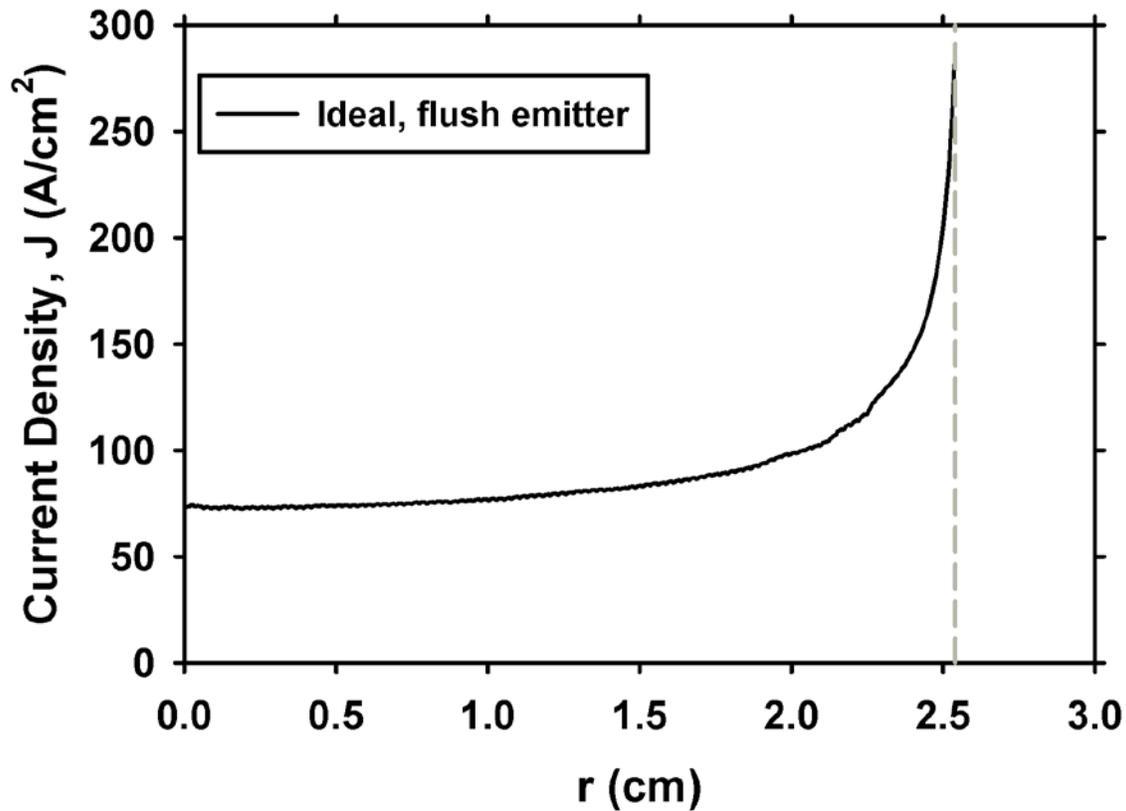

equal currents.)

Figure 6: Current density across the ideal, flush emitter shown in Fig. 5. The total current is 2.1 kA with $V_{AK}$=3.8 MV. The outer edge of the 2-inch diameter cathode is indicated by the dashed line.

The current density as a function of cathode emission radius for this ideal case is shown in Fig. 7. Here it can be seen that, even as the emission area becomes large, the current density never approaches a constant value, which would be the scaling predicted for an infinite planar diode. In fact, as seen in Fig. 8, the current scaling with cathode size is approximately $I \propto R^{1.4}$ rather than the $I \propto R^2$ one would expect for planar diodes with AK gaps much shorter than the cathode size. Moreover, this power-law scaling disagrees with the Langmuir-Blodgett scaling for both spherical [16] and cylindrical diodes [17]. This deviation from analytical geometric-scaling laws of the DARHT Axis-I space-charge limit is largely due to the non-uniform emission (see Fig. 6) resulting from a localized emission region.

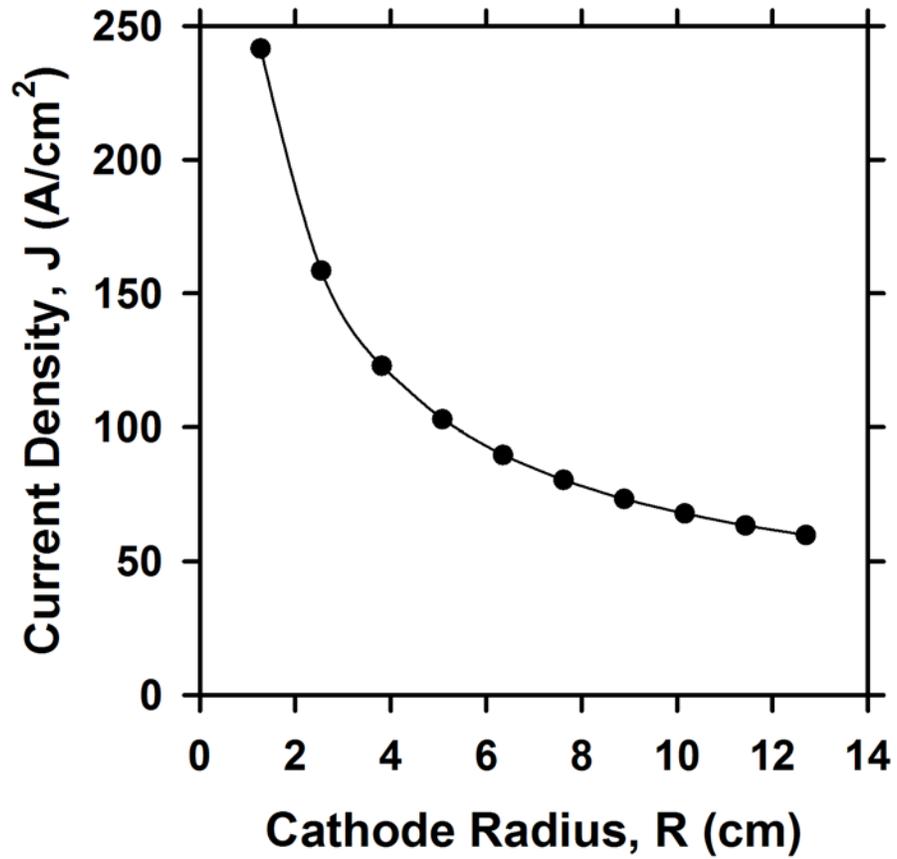

Figure 7: Current density for an ideal flat cathode in the DARHT Axis-I diode as a function of its size. Individual simulation results are shown as filled circles. The connecting line is simply a visual aid, and has no theoretical connotation.

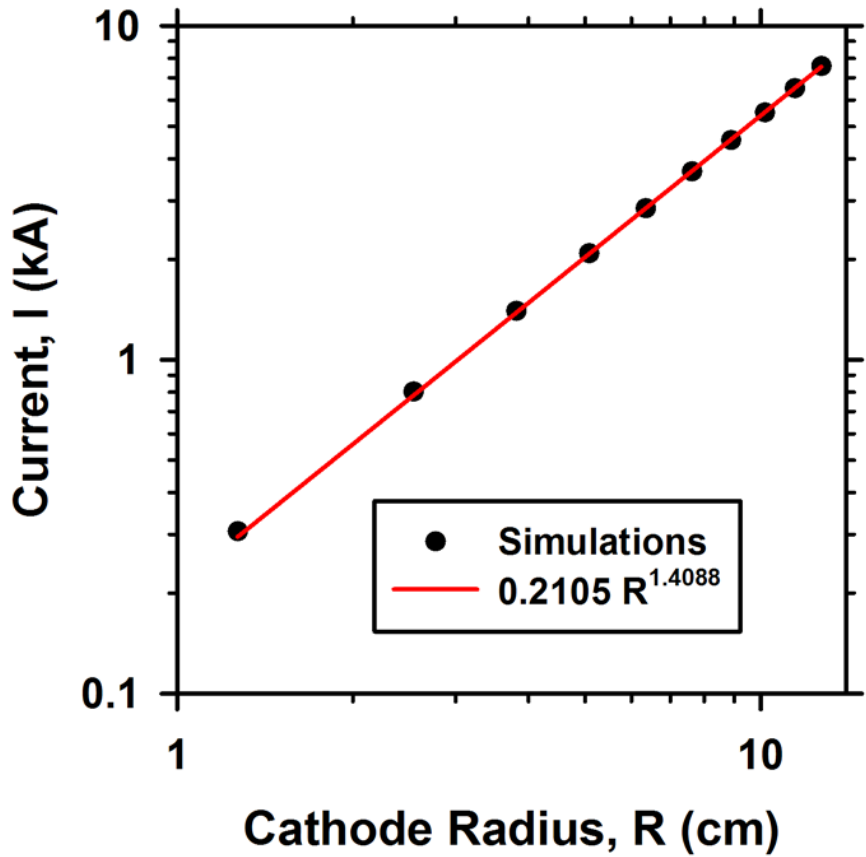

Figure 8: Space-charge limited current for an ideal flat cathode in the DARHT Axis-I diode as a function of its size. Individual simulation results are shown as filled circles. The red line is a least-square power-law fit to the simulation results.

## B. Recessed DARHT Axis-I Cathode

The hollowing of the beam from an ideal flush cathode is exacerbated by the as-built Axis-I cathode, because the as-built cathode has a corner at the edge that really does enhance the field and cause true edge emission (in addition to the space-charge suppression near the center). Figure 9 shows the space-charge limited beam extracted from an as-built cathode surface that is flush. The equipotentials in this figure clearly show the space-charge depression on axis near the emitter, as well as the enhancement at the cathode edge. This simulation produced 2.25 kA with $V_{AK}$=3.8 MV, about 150 A more than the ideal flush emitter studied in the preceding section. The extra current comes from the extra area wrapping around the edge, and the enhancement of the field there producing true edge emission.

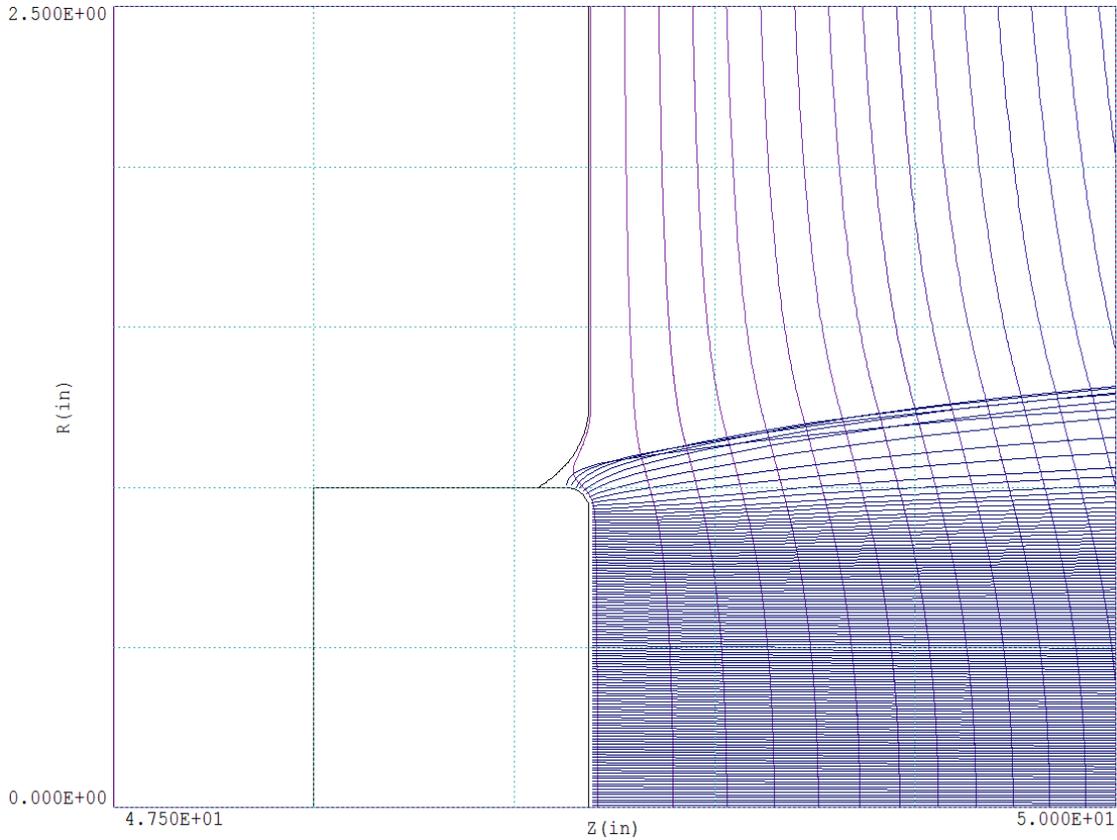

Figure 9: Electron beam extracted from the as-built 2-inch diameter cathode positioned to be flush with the flat surface of the Axis-I cathode shroud. Simulation of this geometry produces 2.25 kA with $V_{AK}$=3.8 MV. (N. B. The density of rays is not equivalent to current density, because the rays do not carry equal currents.)

     As fielded, the Axis-I cathodes are normally retracted below the surface, as shown in Fig. 10. Recessing the cathode has the Pierce-like effect of reducing the emission non-uniformity by flattening the equipotentials in the presence of beam space charge [11,12]. Moreover, retraction significantly reduces the field at the edge, thereby suppressing the real edge-emission. The effect on space-charge limited current density from these effects is clearly shown in Fig. 11, which compares current density distributions for flush (Fig. 9) and retracted (Fig. 10) as-built cathodes.

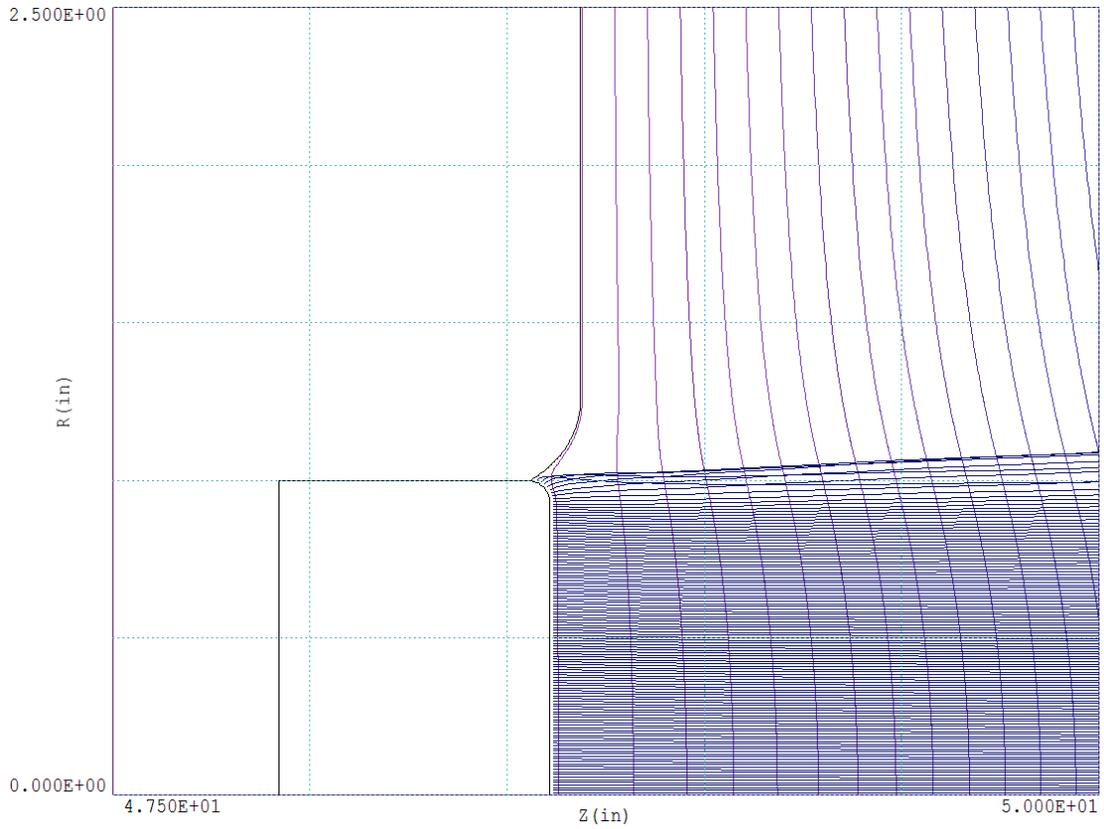

Figure 10: Electron beam extracted from the as-built 2-inch diameter cathode retracted 2 mm below the flat surface of the Axis-I cathode shroud. Simulation of this geometry produces 1.75 kA with $V_{AK}$=3.8 MV. (N. B. The density of rays is not equivalent to current density, because the rays do not carry equal currents.)

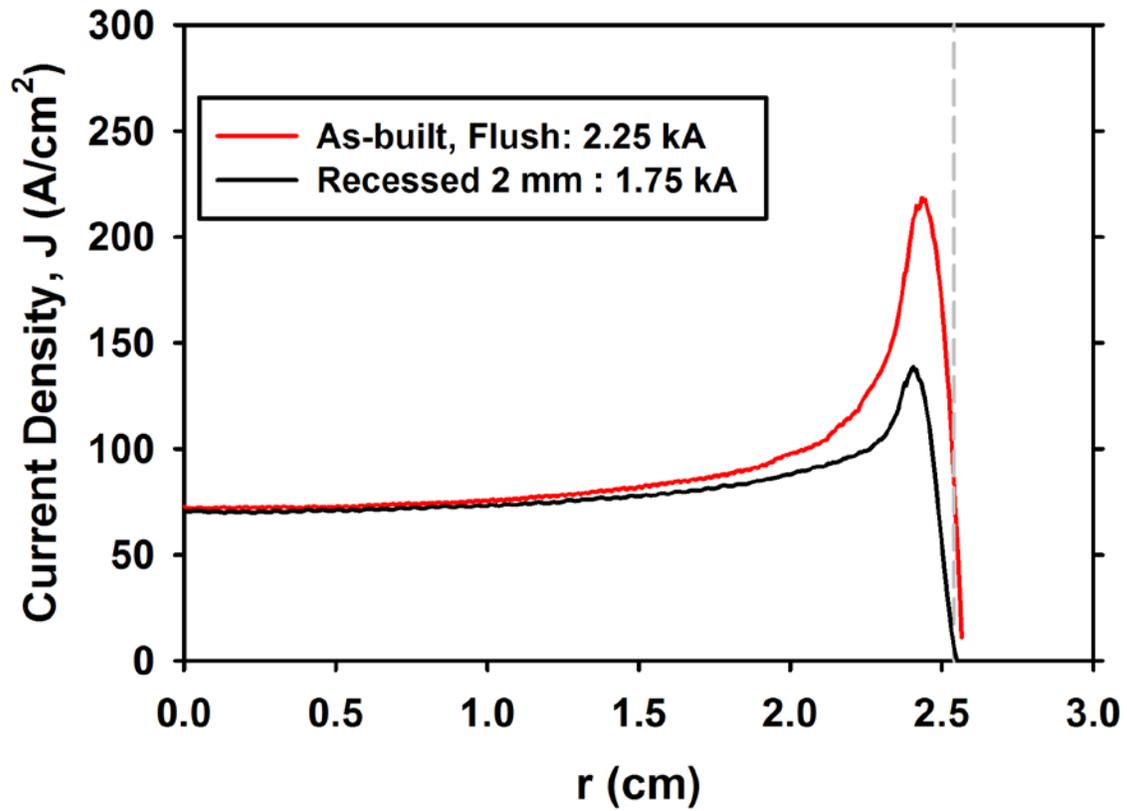

Figure 11: Comparison of current density distributions for as-built cathodes flush with surface (red curve) and retracted by 2 mm (black curve). The outer edge of the 2-inch diameter cathode is shown by the dashed line. The current density is plotted over the entire emission surface, which partially wraps around the cathode edge as shown in Fig. 7 and Fig. 10.

The cathode plasma produced by the explosive emission process rapidly expands [6], so the location of the emission surface is not coincident with the surface of the velvet, and is unknown. Earlier simulations [2] showed that a 2-inch cathode emission surface recessed by 2 mm produced a current close to the 1.74-kA nominal value with 3.8-MV on the diode. Further simulations showed a linear current reduction sensitivity of -252 A/(mm of retraction) as shown in Fig. 12. Because of this uncertainty of the location of the emitting surface, as well as the experimental uncertainty in the actual position of the ends of the velvet tufts, the depth of emission surface is included in the geometrical scaling. For example, the simulated current produced by a 2-inch diameter cathode is 1.74 kA when recessed to an emission depth of ~2 mm. This is about the same as the experimentally observed current from the same diameter cathode when the mounting surface is recessed by 3 mm [1], suggesting that the height of the tufts plus thickness of plasma is ~1 mm..

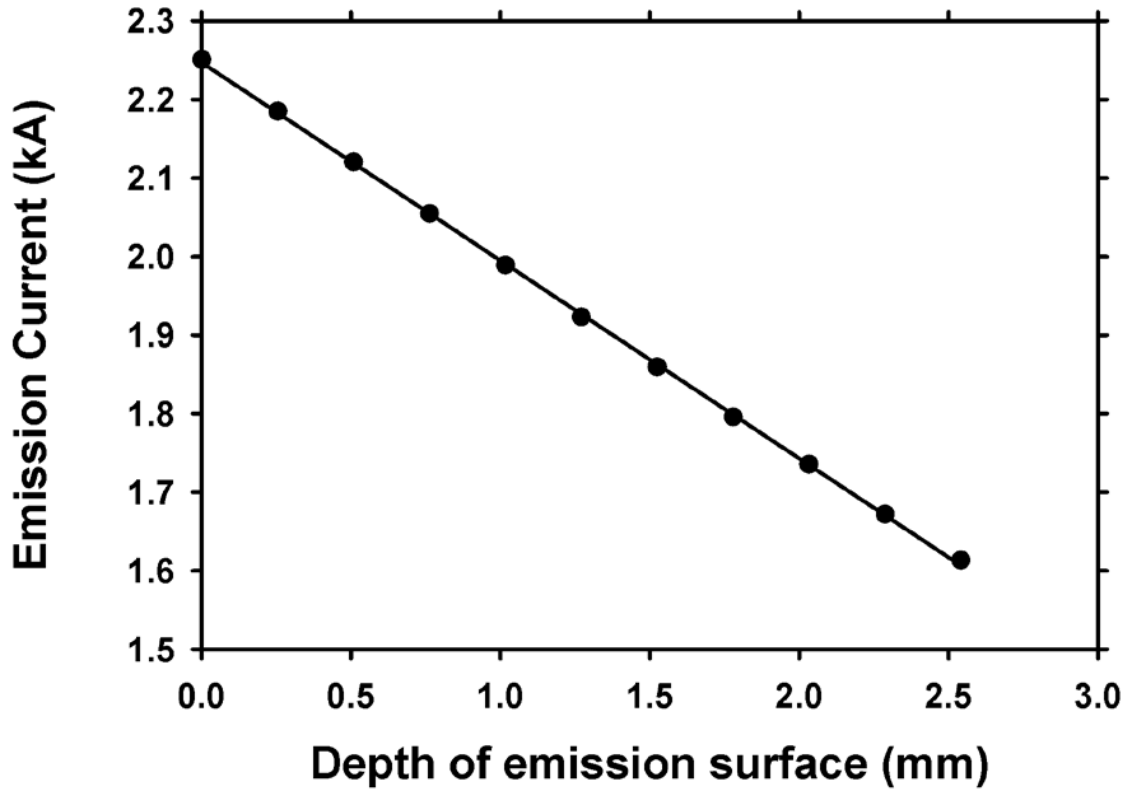

Figure 12: Simulated current emitted by the model of the as-built 2-inch diameter cathode as a function of the depth of retraction below the shroud surface. The line is a least square fit to the simulation results, and it shows the sensitivity to retraction of -252 A/mm.

## C. Nominal Axis-I Cathode Scaling

The DARHT Axis-I pulsewidth is fixed, so one way to adjust the radiographic dose without introducing unwanted attenuation is to change the current. This can be done by using cathodes with different radii and recess depths. The baseline cathodes considered for radiography operations have diameters of 1 inch, 2 inches, and 2.75 inches, an all are recessed to ~2 mm, a dimension with some uncertainty due to velvet cloth thickness variations and plasma expansion. All simulations for these dimensions were performed with an AK voltage of 3.8 MV. Simulated space-charge limited current results are plotted in Fig. 13. Also shown on Fig. 13 is the dose at one meter that could be expected, based on a simple linear-with-current scaling from the nominal 550 Rad(Pt) at 1.75 kA [1] for the fixed pulse width. This illustrates the large range (10:1) of radiograph dose accessible by simply changing cathode size and depth of recess.

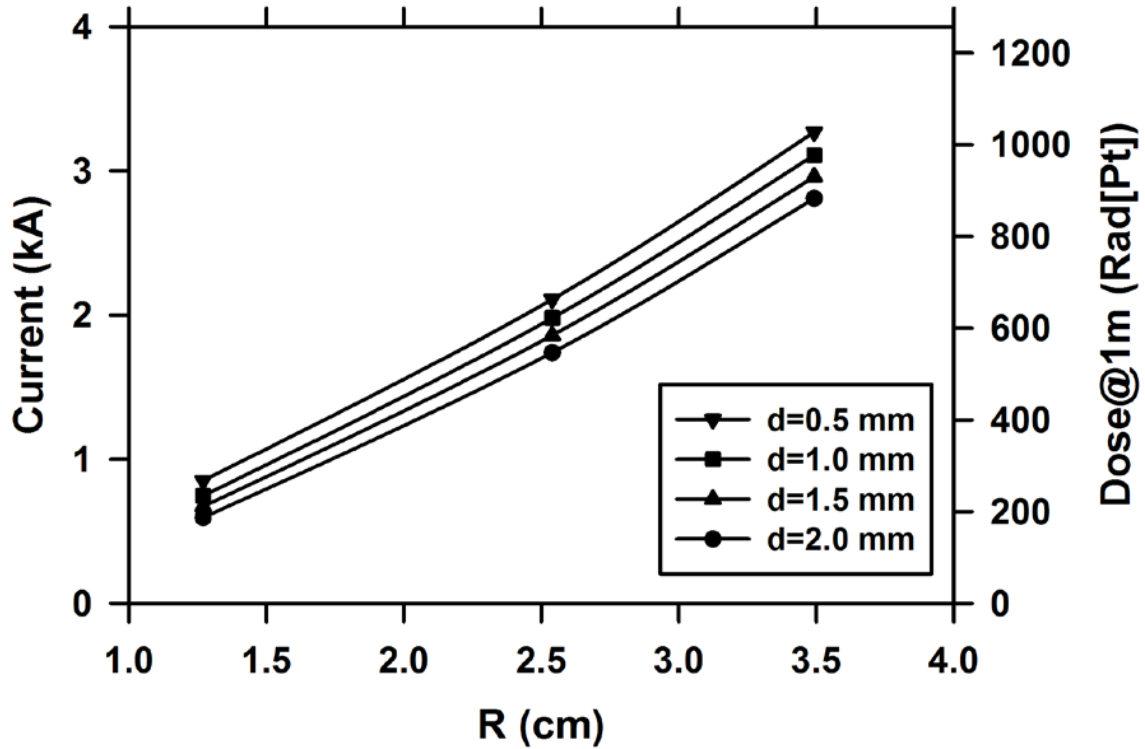

Figure 13: Space-charge limited current as a function of cathode radius $R$ and depth of the emission surface $d$.

## III. Discussion

Rule-of-thumb scaling laws for experimental planning can be derived from these simulations. Figure 14 shows the data re-plotted on logarithmic axes. Here, it is evident that the scaling of current (or dose) has a power-law relation to cathode size, and that the power is dependent on the recess depth, as is the constant multiplier. That is, the current is given by $I(R,d:3.8\text{MV}) = G(d)R^{k(d)}$. The best fits to the simulations give

$$I(R,d:3.8\text{MV}) = G(d)R^{k(d)}$$
$$G(d) = 0.664 - 0.125 d_{mm} \text{ kA} \quad (1)$$
$$k(d) = 1.29 + 0.118 d_{mm}$$

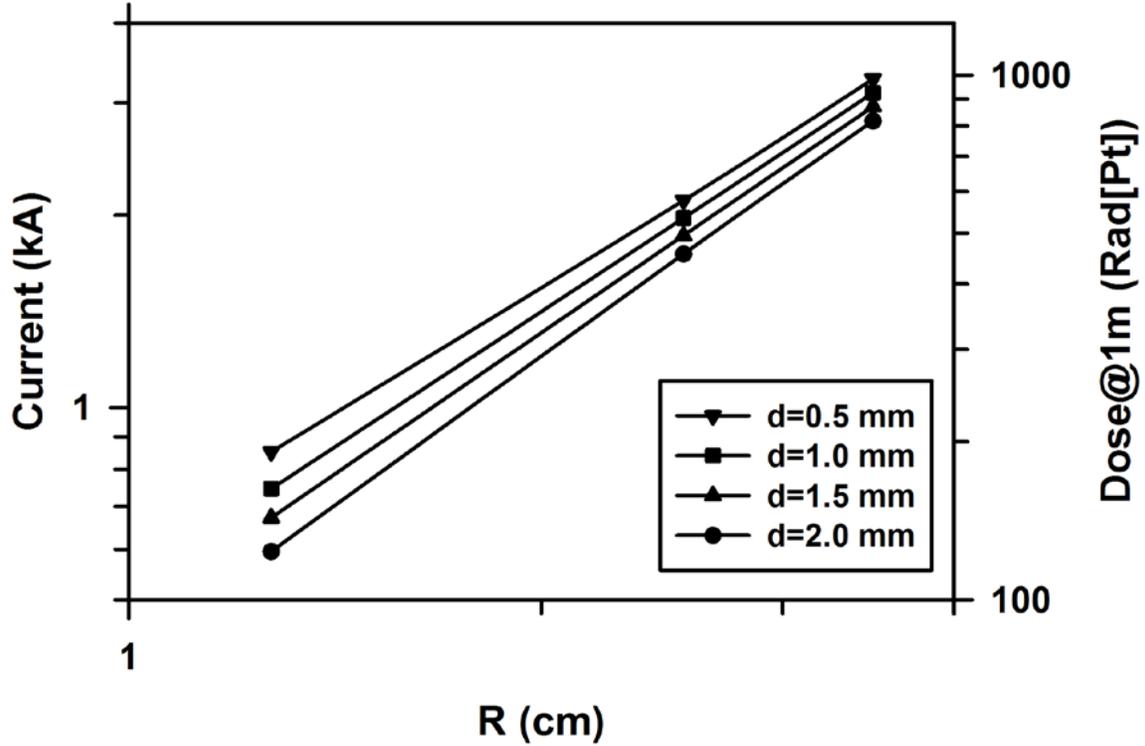

Figure 14: Space-charge limited current as a function of cathode radius $R$ and depth of the emission surface $d$.

Finally, for experimental planning purposes it is useful to have a simple rule-of-thumb for the required radius as a function of the desired current. For the 2-mm emission surface depth corresponding to the nominal 3-mm depth of the backing plate this is approximately

$$R \approx [I/0.419]^{2/3} = [D/132]^{2/3} \quad (2)$$

where $I$ is in kA, $R$ is in cm, and $D$ is in Rad at one meter. This is almost exact for the standard 2-inch cathode, but may not be so accurate for smaller sizes, because of significant differences in edge emission.

## IV. Conclusions

Simulations of the DARHT Axis-I diode validate that a 10:1 range of current and radiographic dose can be achieved by changing the size of the cathode. This provides a means for adjusting the dose to meet experimental needs without introducing detrimental attenuation[1]. In particular, a 2.75-inch diameter cathode, with the backing plate recessed the nominal 3 mm below the flat shroud surface, is predicted to produce a 2.81-kA beam. A simple scaling law that fits the simulation results when the cathode backing plate is recessed 3-mm below the surface is $I = 0.419 R^{1.52}$, where I is in kA and R is in cm. This can be inverted to give a design rule-of-thumb $R \approx [I/0.419]^{2/3} = [D/132]^{2/3}$ for choosing cathode sizes based on dose required at one meter.

## Acknowledgements

The author acknowledges stimulating discussions with my WX-5 colleagues, in particular Dave Moir and Trent McCuistian, who are the principal experimenters on the DARHT Axis-I injector. This work was supported by the US National Nuclear Security Agency and the US Department of Energy under contract W-7405-ENG-36.